\documentclass[twocolumn, superscriptaddress, secnumarabic, amssymb, showpacs, nobibnotes, aps, prb]{revtex4-2}
\usepackage{graphicx }
\usepackage{dcolumn}
\usepackage{bm}
\usepackage{xcolor}
\usepackage{upgreek}
\usepackage{xspace}
\usepackage{amsmath}
\usepackage[]{hyperref}
\hypersetup{colorlinks=true,linkcolor=blue,citecolor=blue,urlcolor=blue,pdfpagemode=UseNone}

\begin{document}
\newcommand{\pbcute}{PbCuTe$_2$O$_6$}
\newcommand{\muSR}{$\mu$SR\xspace}
\newcommand{\mus}{$\upmu$s\textsuperscript{-1}\xspace}
\newcommand{\REF}{{\bf[REF]}}

\title{Lifting the degeneracy of 
quantum spin liquid phase by uniaxial pressure}
\author{Shams Sohel Islam}
\email{shams.islam@psi.ch}
\author{Zurab Guguchia}
\author{Orion Gerguri}
\author{Petr Kr\'al}
\author{Maxime Lamotte}
\author{Toni Shiroka}
\author{Jonas A. Krieger}
\author{Thomas J. Hicken}
\author{Tina Arh}
\author{Gediminas Simutis}
\affiliation{PSI Center for Neutron and Muon Sciences CNM, 5232 Villigen PSI, Switzerland}
\author{Abanoub Hanna} 
\author{Nazmul Islam}
\author{Bella Lake}
\affiliation{Helmholtz-Zentrum Berlin f\"ur Materialien und Energie GmbH, Hahn-Meitner-Platz~1, D-14109 Berlin, Germany}
\author{Hubertus Luetkens}
\affiliation{PSI Center for Neutron and Muon Sciences CNM, 5232 Villigen PSI, Switzerland}
\author{Hans Henning Klauss}
\affiliation{Institute of Solid State
and Materials Physics, TU Dresden, D-01062 Dresden, Germany}
\author{Rajib Sarkar}
\email{rajib.sarkar@tu-dresden.de}
\affiliation{Institute of Solid State
and Materials Physics, TU Dresden, D-01062 Dresden, Germany}

\date{\today}

\begin{abstract}
We report muon spin relaxation/rotation ($\mu$SR) measurements of the candidate three-dimensional (3D) quantum spin liquid (QSL) PbCuTe$_2$O$_6$, hosting $S=1/2$ moments, under controlled in situ [110] uniaxial compression up to $\sigma_{[110]}=37.7$~MPa. A small directional lattice perturbation significantly modifies the local magnetic response, while above $\sigma_{\rm cr}\sim10.8$\,MPa the relaxation rates are strongly enhanced and the internal-field distribution is substantially broadened.
These changes occur along  with the local crystalline symmetry breaking. While, no evidence for conventional static long-range magnetic order is observed, the compression drives the system towards a structurally modified and strongly correlated state in which enhanced quasi-static correlations coexist with persistent slow spin dynamics. This work demonstrates a clean and symmetry-selective route to control frustrated exchange landscape and access hidden magnetic instabilities in a 3D QSL candidate opening up the possibilities to tune other correlated systems where intrinsic coupling between magnetic and lattice degrees of freedom are relevant.
\end{abstract}
\maketitle
\newpage
\noindent
Quantum spin liquids (QSLs) are novel states of matter in which electron spins remain fluctuating yet highly entangled, without developing conventional magnetic long-range order (LRO) down to absolute zero temperatures. Such states can host fractionalized excitations, such as spinons, and are often associated with emergent gauge structures and nontrivial topology~\cite{Broholmeaay0668,Knolle451}. Despite decades of effort, clean experimental realizations remain scarce, particularly in three-dimensional (3D) magnets, where enhanced connectivity and reduced quantum fluctuations generally favor magnetically ordered ground states~\cite{Gingras056501,Shockley047201,Balz942,Zivkovic157204,Gao1052}.

PbCuTe$_2$O$_6$ represents a rare 3D $S=1/2$ QSL candidate. As shown in Fig.~1(a), competing antiferromagnetic interactions between Cu$^{2+}$ spins form a frustrated 3D hyper-hyperkagome network governed by nearly equal dominant $J_1$ and $J_2$ couplings, with subdominant $J_3$ and weaker $J_4$ interactions providing additional chainlike magnetic connections~\cite{chillal2348, Kote035141}. Evidence for QSL behavior includes a broad spinon-like magnetic continuum observed by inelastic neutron scattering, persistent spin fluctuations down to 20~mK revealed by ZF-$\mu$SR, and a sizeable linear-in-$T$ thermal-conductivity contribution consistent with itinerant spinon heat transport~\cite{chillal2348,Khuntia107203,Hong256701}.

Another crucial feature of PbCuTe$_2$O$_6$ is the strong coupling between magnetic, lattice, and dielectric degrees of freedoms at low temperatures. Thermodynamic studies on single crystals revealed a ferroelectric transition near $T_{\mathrm{FE}}\simeq1$~K, accompanied by anisotropic lattice distortions and a modified magnetic response that remains consistent with a potential QSL state~\cite{Thurn95,Eibisch235133,Hanna113401}. Direction-resolved thermal-expansion and dielectric measurements show the strongest responses along [110], while the divergence of the magnetic Gr\"uneisen parameter indicates proximity to quantum criticality. These observations suggest that the proposed QSL state is particularly sensitive to lattice distortions along the [110] direction.

Uniaxial strain provides a clean route to apply controlled perturbations to frustrated magnets without introducing chemical disorder. Unlike hydrostatic pressure, which mainly tunes overall exchange energy scales~\cite{Chatterjee136701,Shimizu107203}, uniaxial strain imposes a directional lattice distortion that can selectively modify exchange pathways and lattice symmetry~\cite{Islam2025,Guguchia097005,Guguchiae2303423120,Grinenko748}. Yet, controlled strain tuning of frustrated magnetism remains experimentally limited, with studies focused primarily on quasi-two-dimensional systems~\cite{Liebericheadz0699,Wang256501}. For PbCuTe$_2$O$_6$, the pronounced [110] sensitivity provides a direct way to probe the microscopic balance of interactions that stabilizes the QSL-like state in a three-dimensional magnet. This raises a central question: does [110] compression release frustration and drive the system toward magnetic order, or does the 3D QSL-like state remain dynamically fluctuating despite the symmetry-breaking perturbation? Addressing this question requires a local magnetic probe capable of tracking static, quasi-static, and dynamic correlations under controlled strain.

\begin{figure}[hbt]
\includegraphics[width=\columnwidth]{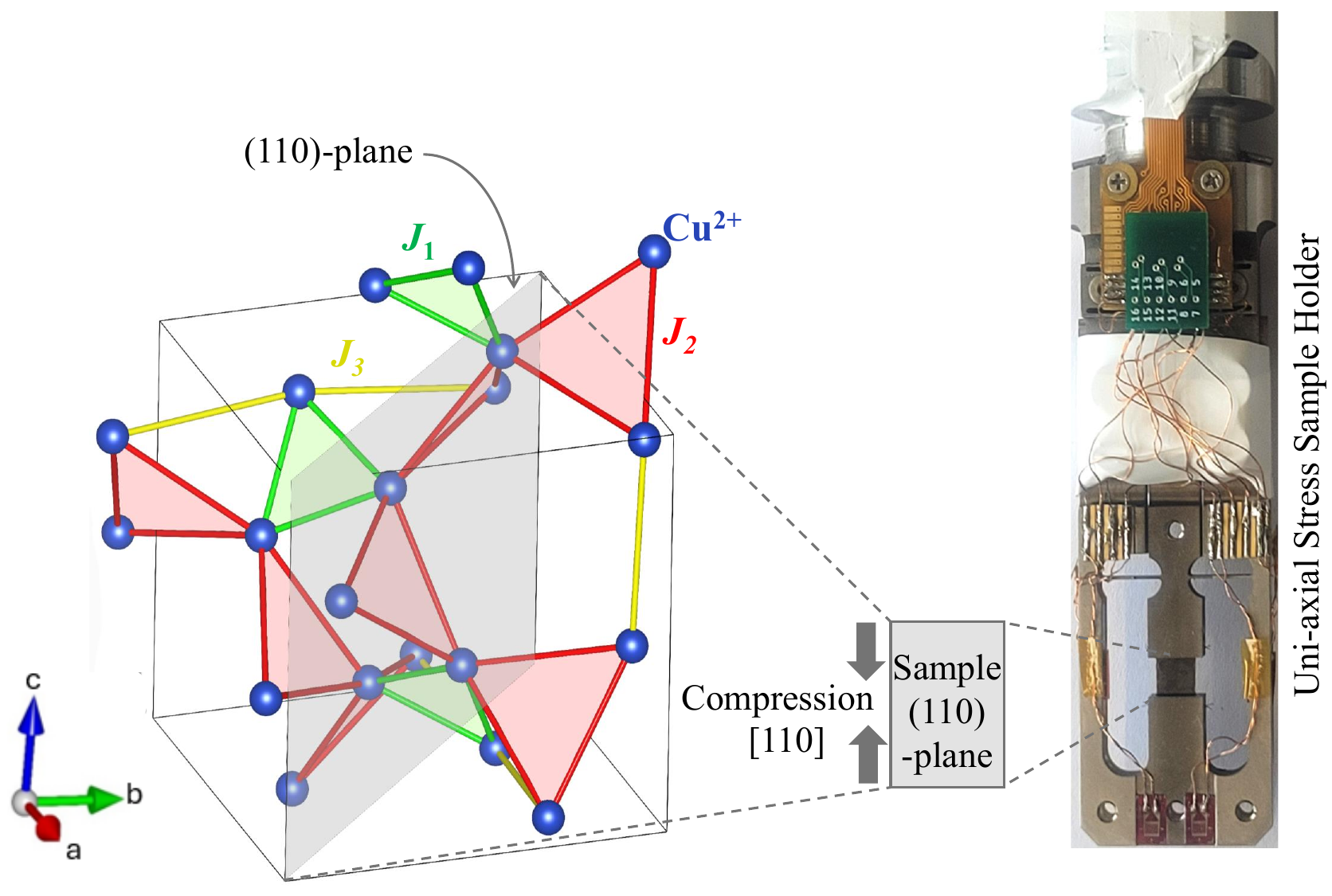}
\caption{\label{fig:crystal-structure}
(a) Schematic of the 3D frustrated Cu$^{2+}$ network in PbCuTe$_2$O$_6$, highlighting representative competing exchange pathways $J_1$, $J_2$, and $J_3$ and the $(110)$ plane relevant to the uniaxial-stress geometry. The arrow indicates the applied compression direction, [110]. For clarity, the $J_4$ interaction, which forms an additional chain-like coupling along the body diagonal, is not shown~\cite{Fancelli184415}. (b) Photograph of the sample mounted on the uniaxial-stress sample holder for $\mu$SR.}
\end{figure}

Here, we use $\mu$SR to microscopically investigate PbCuTe$_2$O$_6$ under in-situ uniaxial compression along [110] using a piezoelectric-driven stress device as shown in Fig.~\ref{fig:crystal-structure}(b)~\cite{Hicks2018,Ghosh2020}. We show that a small [110] uniaxial stress modifies both the local crystalline environment and the magnetic response, revealing a stress-induced reconstruction of the low-energy magnetic state. These results establish uniaxial strain as a clean route to tune 3D QSLs, directly exploiting magnetoelastic coupling to reshape the frustrated magnetic landscape.

Using zero-field (ZF), transverse-field (TF), and longitudinal-field (LF) $\mu$SR, we investigated the magnetic response of the same PbCuTe$_2$O$_6$ single crystal, first at ambient pressure and then under in-situ [110] uniaxial compression after mounting in the strain device. All asymmetry spectra are background-subtracted and normalized to the maximum observable relaxing signal; details are provided in the Supplemental Material (SM)~\cite{SM}. Implanted muons are highly sensitive to local fields from neighboring Cu$^{2+}$ spins, with a field resolution of $\approx0.1$~mT, making $\mu$SR well suited for detecting weak static magnetism~\cite{Hillier2022}. Its sensitivity to spin fluctuations over a broad timescale further allows one to distinguish static, quasi-static, and dynamic magnetic responses.

\begin{figure*}[hbt]
\includegraphics[width=\textwidth]{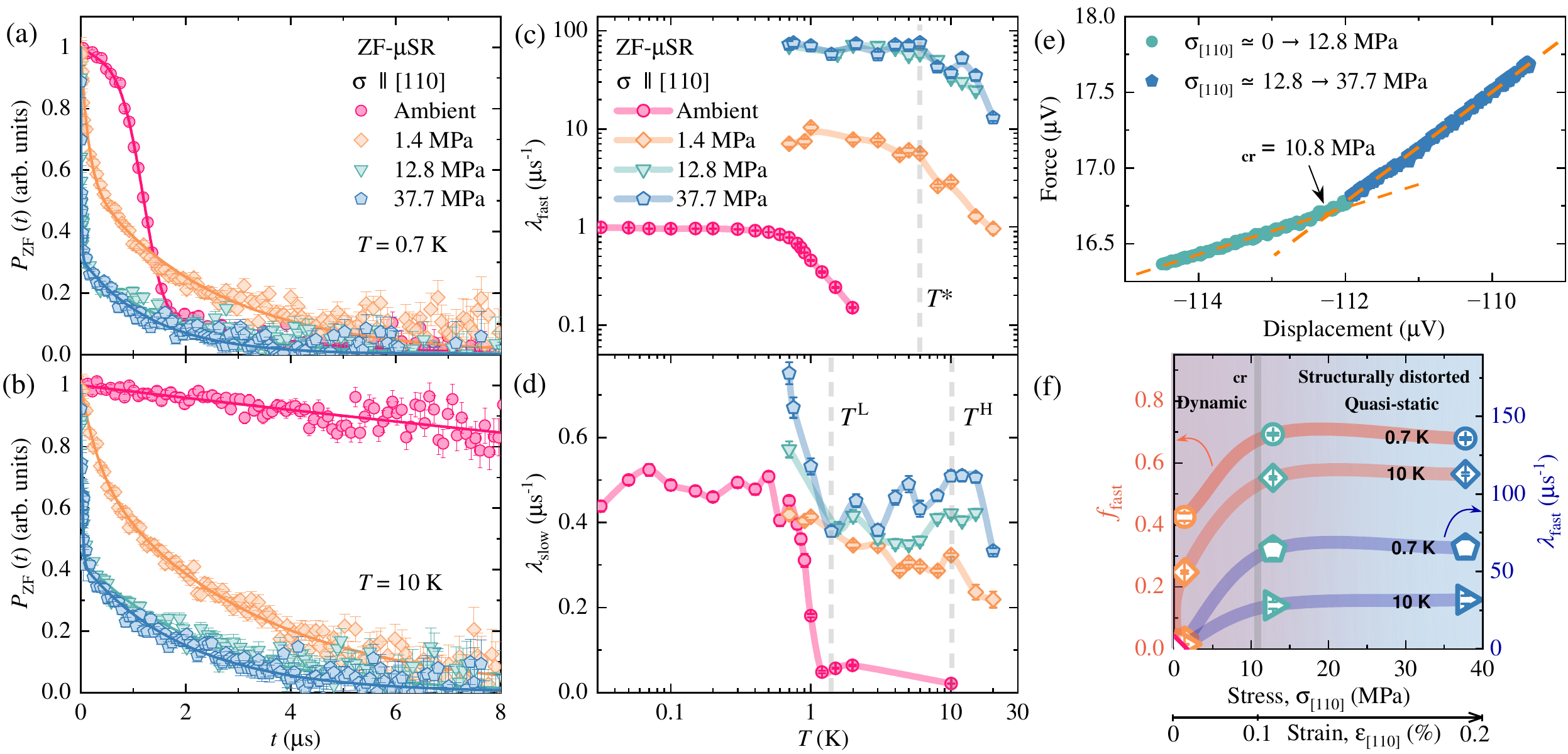}  
\caption{\label{fig:ZF-musr-spectra} (a),(b) Representative ZF-\(\mu\)SR depolarization measured as a function of uniaxial stress \(\sigma_{[110]}\) at \(T=0.7~\mathrm{K}\) and \(10~\mathrm{K}\), respectively. At ambient pressure, these temperatures correspond to the proposed QSL-like regime and a regime closer to the paramagnetic state, respectively~\cite{Khuntia107203}. Solid lines are fits using the two-component relaxation model described in the SM~\cite{SM}.
(c),(d) Temperature dependence of the fast and slow ZF relaxation rates for selected uniaxial stresses. Vertical dashed lines indicate the characteristic temperature scales discussed in the text. 
(e) Stress--strain curve: force versus displacement measured during the uniaxial-stress experiment. The arrow marks the slope change at the critical stress $\sigma_{\mathrm{cr}}$. 
(f) Stress dependence of the fast-relaxing fraction $f_{\mathrm{fast}}$ and relaxation rate $\lambda_{\mathrm{fast}}$. The vertical solid grey line marks $\sigma_{\mathrm{cr}}$.}
\end{figure*}

Figures~\ref{fig:ZF-musr-spectra}(a) and (b) show representative ZF-$\mu$SR spectra measured at 0.7~K and 10~K under ambient pressure and selected uniaxial stresses; the full temperature dependence is provided in the SM~\cite{SM}. With increasing $\sigma_{[110]}$, the spectra show a progressive enhancement of the early-time depolarization at both temperatures, demonstrating that [110] compression strongly modifies the local magnetic response over an extended temperature range. No coherent oscillations are observed at any stress, indicating the absence of a well-defined static internal field associated with conventional magnetic LRO down to 0.7~K. At 37.7~MPa, the rapid initial depolarization is followed by a slowly relaxing long-time component, showing that slow spin dynamics persist even at the highest stress. To quantify this stress-dependent temperature evolution, the spectra were fitted using a phenomenological two-component relaxation model with fast and slow relaxation rates, $\lambda_{\mathrm{fast}}$ and $\lambda_{\mathrm{slow}}$, shown in Figs.~\ref{fig:ZF-musr-spectra}(c) and (d); details are provided in the SM~\cite{SM}.

At ambient pressure, $\lambda_{\mathrm{fast}}$ increases below $\sim2$~K, indicating slowing spin dynamics and the development of low-frequency magnetic correlations, while $\lambda_{\mathrm{slow}}$ shows a stronger enhancement below $\sim1$~K. Below $\sim0.7$~K, both relaxation rates tend to saturate, consistent with persistent low-energy spin fluctuations characteristic of a dynamical QSL-like ground state~\cite{Khuntia107203,Zivkovic157204}. These temperature scales coincide with thermodynamic and dielectric anomalies, where anisotropic lattice distortions indicative of deviations from cubic symmetry develop below $\sim2$~K and ferroelectric order appears at $\sim1$~K~\cite{Thurn95,Eibisch235133}. The emergence of two-component relaxation below $\sim2$~K therefore suggests spatially inhomogeneous relaxation, possibly arising from coexisting undistorted and distorted local regions with distinct spin dynamics; a detailed discussion is provided in the SM~\cite{SM}.

Under uniaxial stress, the relaxation rates change dramatically. The fast relaxation rate $\lambda_{\mathrm{fast}}$ increases by nearly two orders of magnitude from ambient pressure to 37.7~MPa, reaching $\sim75~\mu\mathrm{s}^{-1}$ at low temperatures. In contrast, $\lambda_{\mathrm{slow}}$ evolves more gradually and remains much less stress-sensitive, staying comparable to the ambient-pressure value even at 37.7~MPa. This could suggest that the underlying spin-fluctuation rate ($\nu$) is only weakly affected by stress, while the dramatic enhancement of $\lambda_{\mathrm{fast}}$ mainly reflects a broadening of the local-field distribution width ($\Delta$) as the system develops enhanced quasi-static magnetic correlations under increasing stress.

The temperature dependence further reveals multiple characteristic energy scales, as shown in Figs.~\ref{fig:ZF-musr-spectra}(c) and (d). The onset of the low-temperature plateau in $\lambda_{\mathrm{fast}}$ shifts from $\sim0.7$~K at ambient pressure to $T^{*}\simeq6$~K under uniaxial stress and remains nearly unchanged for the two highest applied stresses. In contrast to the ambient-pressure response, $\lambda_{\mathrm{slow}}$ exhibits a broad high-temperature feature around $\sim10$~K ($T^{\rm H}$), together with a stronger enhancement at low temperatures below $\sim1.7$~K ($T^{\rm L}$), both of which become increasingly pronounced under stress. Taken together, the ZF-\(\mu\)SR results show that [110] compression profoundly alters the low-temperature magnetic behavior of PbCuTe$_2$O$_6$, shifting the dominant relaxation scale from sub-kelvin temperatures to the several-kelvin regime and producing additional stress-enhanced crossover features.

Next, we examine whether the stress-induced magnetic response is accompanied by any structural change. Figure~\ref{fig:ZF-musr-spectra}(e) shows the force--displacement curve measured during the uniaxial-stress experiment. In such measurements, a linear force--displacement response is expected in the absence of structural transformation or plastic deformation, whereas a change in slope can indicate a modification of the sample's mechanical response~\cite{Islam2025,Guguchia097005,Guguchiae2303423120}. Here, the curve exhibits a clear slope change at $\sigma_{\mathrm{cr}}=10.8$~MPa, providing indirect evidence for a stress-induced structural modification.

Importantly, $\sigma_{\mathrm{cr}}$ closely correlates with the ZF-$\mu$SR response. As shown in Fig.~\ref{fig:ZF-musr-spectra}(f), both the fast-relaxing fraction $f_{\mathrm{fast}}$ and the fast relaxation rate $\lambda_{\mathrm{fast}}$ increase rapidly with applied stress before saturating above $\sigma_{\mathrm{cr}}$ at both 0.7~K and 10~K. Moreover, the temperature dependences of $\lambda_{\mathrm{fast}}$ measured at 12.8 and 37.7~MPa nearly overlap. These observations suggest that the applied stress progressively modifies the ambient-pressure crystal structure, until a fully developed stress-induced structurally distorted state is stabilized above $\sigma_{\mathrm{cr}}$. This distorted state then gives rise to a common magnetic response for all stresses above $\sigma_{\mathrm{cr}}$. Consistently, after releasing the stress to the zero-force condition, the relaxation response remains close to that measured at 37.7~MPa rather than recovering the initial ambient-pressure behavior [see SM~\cite{SM}], indicating a largely irreversible stress-induced structural modification.

\begin{figure*}[hbt]
\includegraphics[width=\linewidth]{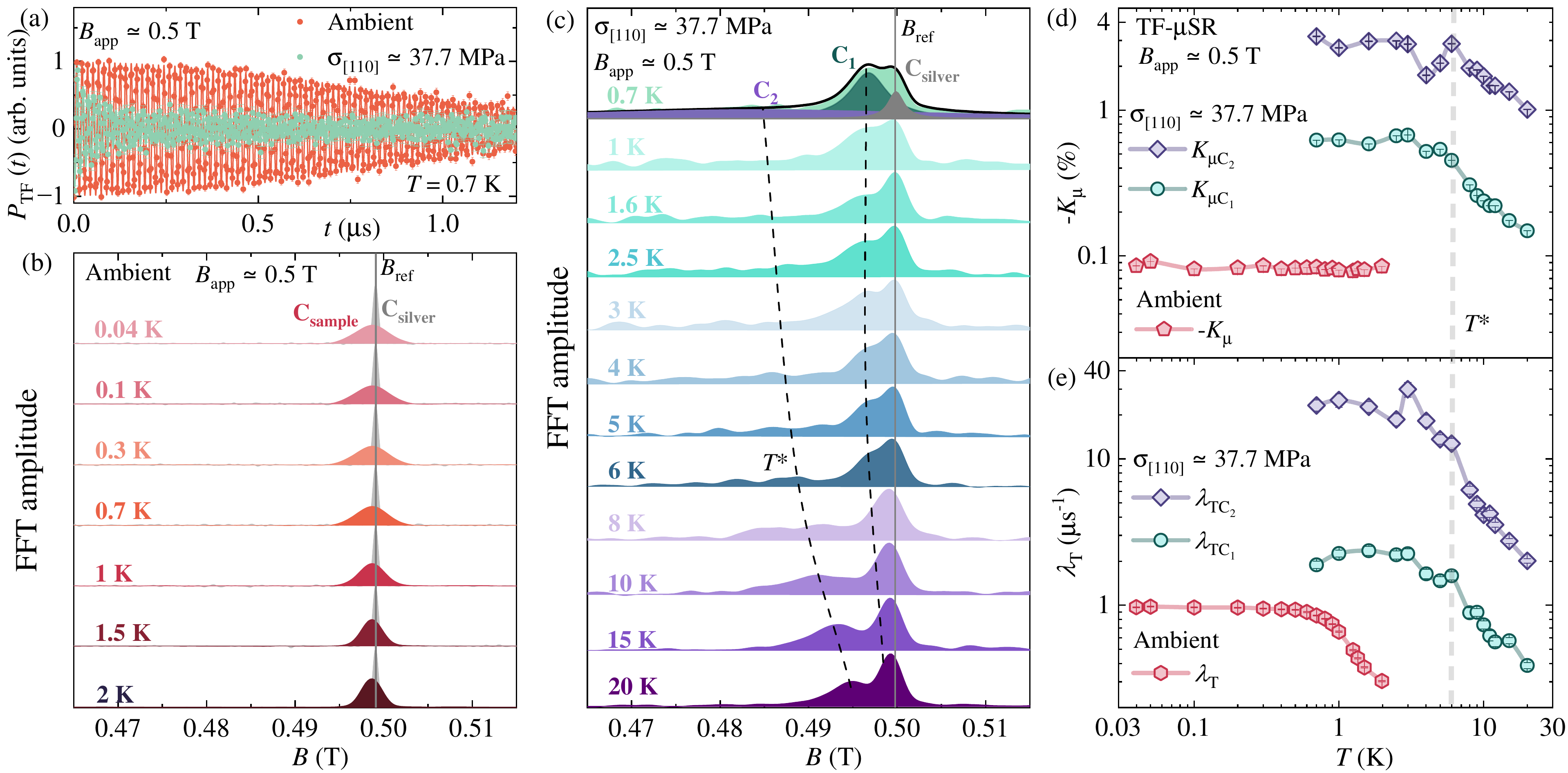}
\caption{\label{fig:knightshift}
(a) Representative TF-\(\mu\)SR depolarization measured at \(T=0.7~\mathrm{K}\) under ambient pressure and \(\sigma_{[110]}=37.7~\mathrm{MPa}\), in an applied transverse field of \(B_{\rm app}\simeq0.5~\mathrm{T}\). The solid lines represent fits described in the SM~\cite{SM}.
(b),(c) Fourier-transform spectra measured at representative temperatures under ambient pressure and \(\sigma_{[110]}=37.7~\mathrm{MPa}\), respectively, in the same applied field. At \(\sigma_{[110]}=37.7~\mathrm{MPa}\), the decomposition of the \(0.7~\mathrm{K}\) spectra into two sample-related peaks (C$_1$ and C$_2$) and a reference peak (C$_{\rm silver}$) is shown. The vertical solid line marks the reference field determined from the silver signal.
(d) Temperature dependence of the muon Knight shifts under ambient pressure and \(\sigma_{[110]}=37.7~\mathrm{MPa}\).
(e) Temperature dependence of the TF relaxation rates under ambient pressure and \(\sigma_{[110]}=37.7~\mathrm{MPa}\).
}
\end{figure*}

We next use TF-$\mu$SR to probe the local-field distribution under uniaxial stress. In this geometry, the muon precession frequency measures the average local field (vector sum of external and internal fields) at the muon site, while the damping reflects dephasing arising from the distribution of local fields. Figure~\ref{fig:knightshift}(a) shows representative spectra measured in $B_{\rm app}=0.5$~T at $T=0.7$~K under ambient pressure and at $\sigma_{[110]}=37.7$~MPa; additional TF-$\mu$SR spectra over the full temperature range and fitting details are provided in the SM~\cite{SM}. At ambient pressure, the long-lived oscillations indicate a relatively narrow field distribution, whereas under stress the strongly damped oscillations demonstrate substantial broadening of the local-field distribution.

This broadening is directly resolved in the FFT spectra shown in Figs.~\ref{fig:knightshift}(b) and (c). At ambient pressure, the spectra are dominated by a single sample component, C$_{\rm sample}$, with weak temperature dependence. In contrast, at 37.7~MPa the field distribution is already broadened at high-$T$ near $\sim20$~K, indicating stress-induced quasi-static local-field broadening over an extended temperature range. Upon cooling, two broad components, C$_1$ and C$_2$, become increasingly resolved. Below $T^{*}\sim6$~K, C$_2$ exhibits an abrupt additional broadening while C$_1$ separates further from the silver reference signal. This coincides with the characteristic scale identified in the fast ZF-$\mu$SR relaxation rate and marks a further enhancement of the quasi-static local-field distribution~\cite{Amato_PhysRevB.89.184425, Lancaster_PhysRevB.93.140412}. Furthermore, the emergence of two sample components provides strong evidence that the muons sense two inequivalent local environments under stress. Together with the change in slope of the stress--strain curve, this suggests a stress-induced lowering of the crystalline symmetry~\cite{Guguchia097005, Guguchiae2303423120}. This also explains the pronounced two-component relaxation observed in ZF-\(\mu\)SR under stress.

Figure~\ref{fig:knightshift}(d) shows the Knight shifts extracted from the TF-$\mu$SR spectra~\cite{SM}. Since the muon Knight shift provides a measure of local magnetic susceptibility, its temperature dependence tracks the evolution of local spin correlations. 
At ambient pressure, $K_{\mu}$ remains finite, nearly temperature independent and does not show an activated decrease at low temperature, consistent with the characteristic of a gapless QSL-like state~\cite{KhuntiaNatPhys,QuilliamPhysRevB.93.214432,Takahashi2019}.
Under stress, $K_{\mu C_{1}}$ and $K_{\mu C_{2}}$ are enhanced relative to the ambient-pressure response and show distinct temperature dependences, indicating two inequivalent local magnetic environments with a growing local susceptibility upon cooling, which eventually saturates below $T^{*}$.

The corresponding TF relaxation rates are shown in Fig.~\ref{fig:knightshift}(e). Under stress, $\lambda_{\rm TC_{1}}$ and $\lambda_{\rm TC_{2}}$ increase strongly upon cooling and show a discernible change in slope at $T^{*}$. The low-temperature saturation of $\lambda_{\rm T}$, without showing any critical divergence associated with static magnetic ordering, indicates that the stress-induced response does not develop into conventional magnetic LRO.

\begin{figure}[hbt]
\includegraphics[width=\columnwidth]{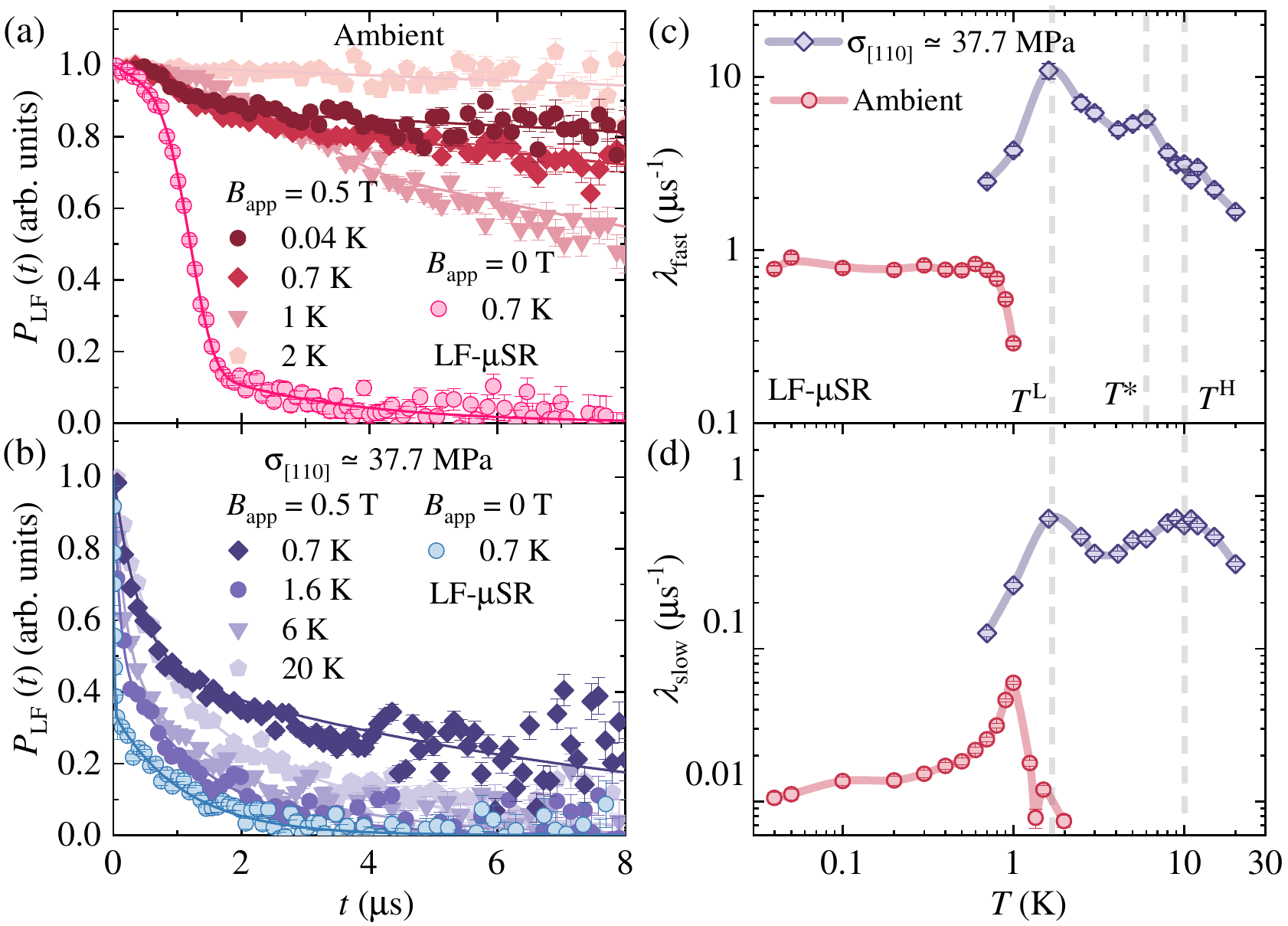}
\caption{\label{fig:LF}(a), (b) Longitudinal-field (LF) $\mu$SR depolarization $P_{\mathrm{LF}}(t)$ measured under ambient pressure and 37.7~MPa, respectively, at selected temperatures in applied fields of $B_{\rm app}=0$ and $0.5$~T. Solid lines are fits using the same two-component relaxation model employed for ZF-$\mu$SR analysis~\cite{SM}. (c), (d) Temperature dependence of the fast ($\lambda_{\mathrm{fast}}$) and slow ($\lambda_{\mathrm{slow}}$) LF relaxation rates under ambient pressure and 37.7~MPa, respectively. Vertical dashed lines mark the characteristic temperature scales discussed in the text.}
\end{figure}

Longitudinal-field (LF) $\mu$SR provides a sensitive means of distinguishing static from dynamic spin relaxation. Figures~\ref{fig:LF}(a) and (b) compare the ZF and LF spectra measured at $0.7$~K for $B_{\rm app}=0$ and $0.5$~T, and show the temperature evolution at a fixed longitudinal field of $B_{\rm app}=0.5$~T under ambient pressure and at the maximum applied stress of $37.7$~MPa, respectively. Detailed LF-data analysis and model comparisons are provided in the SM~\cite{SM}. At ambient pressure, substantial relaxation persists even in $0.5$~T. If the ZF fast relaxation rate $\lambda_{\mathrm{fast}}^{\mathrm{ZF}}\sim1~\mu\mathrm{s}^{-1}$ at $0.7$~K originated solely from static local fields, it would correspond to only $\Delta_{\mu}^{\mathrm{s}}\sim7$~mT, for which a field of order $\sim10\Delta_{\mu}^{\mathrm{s}}$ should nearly decouple the relaxation. The incomplete decoupling therefore indicates the dynamical nature of the ambient-pressure state, as further supported by the dynamic gaussian Kubo--Toyabe analysis presented in the SM~\cite{SM}. In contrast, at $37.7$~MPa, $\lambda_{\mathrm{fast}}^{\mathrm{ZF}}\sim75~\mu\mathrm{s}^{-1}$ corresponds to $\Delta_{\mu}^{\mathrm{s}}\sim0.55$~T, implying that full decoupling would require fields of several tesla, beyond the present experimental range. The Kubo--Toyabe analysis further shows that $\Delta/\nu$ increases by a factor of approximately $17$ under uniaxial stress, indicating a pronounced shift toward the quasistatic regime~\cite{SM}. Nevertheless, comparison with the dynamic exponential Kubo--Toyabe prediction indicates that a purely static picture remains insufficient to describe the magnetic state at the highest applied uniaxial stress of $37.7$~MPa.

Figures~\ref{fig:LF}(c) and (d) show the temperature dependence of the LF relaxation rates extracted using the same two-component model as in the ZF analysis. At ambient pressure, the LF spectra evolve only modestly with temperature, mainly below $\sim2$~K. Correspondingly, $\lambda_{\mathrm{fast}}$ increases below $\sim1$~K and saturates below $\sim0.7$~K. The $\lambda_{\mathrm{slow}}$ shows a pronounced peak near $T_{\mathrm{FE}}\sim1$~K likely reflecting enhanced low-energy magnetic fluctuations modified by the accompanying lattice distortions~\cite{Thurn95,Eibisch235133,Hanna113401}. Upon further cooling, it levels off below 0.3~K, indicating that slow spin fluctuations persist well below $T_{\mathrm{FE}}$ without developing conventional static long-range order, which remains consistent with QSL behavior.

Under uniaxial stress, the LF muon-spin depolarization shows pronounced temperature evolution over a broad temperature range, with both $\lambda_{\mathrm{fast}}$ and $\lambda_{\mathrm{slow}}$ staying elevated up to $20$~K. $\lambda_{\mathrm{fast}}$ exhibits clear anomalies near $T^{\rm L}$ and $T^{*}$, while $\lambda_{\mathrm{slow}}$ develops broader maxima at $T^{\rm L}$ and $T^{\rm H}$. These energy scales closely match those identified from ZF- and TF-$\mu$SR, providing a consistent picture of the stress-induced evolution of low-energy magnetic correlations. Overall, the LF results show that the stress-induced response cannot be described by a purely static local-field distribution and that slow spin dynamics persist under uniaxial stress.

To assess the strain-tuned magnetic response, we first recall that spin and lattice degrees of freedom are strongly intertwined in PbCuTe$_2$O$_6$~\cite{Thurn95,Eibisch235133}. The proposed QSL state is stabilized by competing exchange interactions: the frustrated triangular motifs associated with $J_1$ and $J_2$ couplings ($J_1/k_{\mathrm{B}} \simeq 13.1~\mathrm{K}$, $J_2/k_{\mathrm{B}} \simeq 12.4~\mathrm{K}$) compete with further-neighbor interactions, including the chainlike $J_3$ and the weaker $J_4$ exchange couplings ($J_3/k_{\mathrm{B}} \simeq 6.85~\mathrm{K}$, $J_4/k_{\mathrm{B}} \simeq 1.39~\mathrm{K}$)~\cite{chillal2348}. Theoretical calculations show that changing the relative balance among these interactions, particularly $J_3$ and $J_4$, can drive the system between magnetically ordered states and highly degenerate ground-state manifolds~\cite{Fancelli184415,JIn054408}. Thus, a stress-induced structural modification can reshape the low-energy magnetic response by altering the balance among exchange pathways that are subdominant at ambient pressure.

This perspective is reinforced by recent uniaxial-stress studies showing that small directional perturbations can reorganize finely balanced electronic, magnetic, and structural orders in correlated materials~\cite{Guguchia097005, Guguchiae2303423120, Islam2025, Kissikov2018}, and in frustrated magnets where uniaxial strain tunes the balance among competing interactions~\cite{Liebericheadz0699,Wang256501}. For PbCuTe$_2$O$_6$, high-field measurements with $B\parallel[110]$ further demonstrate its strong magnetoelastic character: magnetic fields produce pronounced lattice distortions and eventually stabilize a field-induced magnetic order above $\sim11$~T~\cite{Eibisch235133}. In the present work, we use [110] uniaxial strain as a direct lattice-tuning approach to perturb the magnetic exchange balance. 

The pronounced change in the magnetic response already at the smallest applied stress of $1.4$~MPa highlights the extreme sensitivity of PbCuTe$_2$O$_6$ to minute lattice perturbations. Microscopically, the emergence of two inequivalent local muon environments revealed by TF-$\mu$SR data at 37.7~MPa, together with the slope change in the stress--strain curve, points to a stress-induced modification of the local crystalline symmetry above $\sigma_{\mathrm{cr}}$. This suggests that the local crystal structure is progressively modified by applied stress, until a structurally distorted state is stabilized above $\sigma_{\mathrm{cr}}\sim10.8$~MPa, corresponding to a strain of $\epsilon\sim0.1\%$. Such sensitivity is consistent with a finely balanced exchange landscape in which competing magnetic configurations are closely spaced in energy.

The strong enhancement of the ZF/TF relaxation rates and the broadening of the TF FFT spectra show that the local internal-field distribution is strongly broadened under stress. Consistently, the Knight shift exhibits a pronounced temperature dependence under stress, in contrast to the nearly temperature-independent ambient-pressure response, indicating an enhanced local susceptibility. Rather than reflecting a uniform enhancement of a single exchange scale, the stress-induced response reveals a hierarchy of low-energy magnetic correlations. In particular, the appearance of multiple crossover features at higher temperatures indicates the emergence of distinct regimes of short-range magnetic correlations. Interestingly, the characteristic scales $T^{\rm L}$ and $T^{\rm H}$ are comparable to the reported chain-like $J_4$ and triangular $J_2$ exchange energy scales, respectively as discussed above. In addition, the robust $T^{*}\sim6$~K scale, visible across all stress-dependent $\mu$SR data, closely matches the magnetic transition in the isostructural compounds (Sr/Ba)CuTe$_2$O$_6$, where chainlike $J_3$ correlations dominate~\cite{Samartzis184435,Chillal144402,Ahmed214413}.

At the same time, LF-$\mu$SR shows that a purely static description is insufficient, demonstrating that slow spin dynamics persists in the stress-induced state. We therefore interpret the state at the highest applied stress as a structurally modified, strongly correlated regime. Here, the frustrated exchange degeneracy is partially lifted and subdominant exchange channels become more visible. This indicates that the enhanced quasi-static local-field broadening coexists with persistent slow spin dynamics.

In conclusion, $\mu$SR measurements, under a small [110] uniaxial stress, reveal that PbCuTe$_2$O$_6$ evolves from a dynamic QSL state to enhanced quasi-static magnetic correlations. The stress-induced state exhibits a modified local crystalline environment, enhanced local susceptibility, and quasistatic field broadening. However, there is no conclusive evidence for conventional static long-range magnetic order. Instead, [110] compression produces a structurally modified strongly correlated regime by reshaping the exchange-coupling landscape. Here, quasistatic correlations coexist with persistent slow spin dynamics. Future $\mu$SR measurements under uniaxial pressure along other crystallographic axes, combined with diffraction and DFT will shed light into the microscopic mechanism for directional control of the spin--lattice-coupled magnetic landscape in PbCuTe$_2$O$_6$.

Our findings identify directional strain as a clean route to expose hidden magnetic instabilities and control frustrated exchange landscapes in three-dimensional QSLs. This opens up the possibilities also to tune other correlated systems where intrinsic coupling between magnetic and lattice degrees of freedom are relevant. Furthermore, these paves the way to tune a quasi-static and/or static magnetic ordered phase towards a quantum critical instability and even stabilize a fluctuating QSL ground state. 

\textit{Acknowledgments.} This work was financially supported by the Deutsche Forschungsgemeinschaft (DFG) within the SFB 1143 “Correlated Magnetism – From Frustration to Topology”, project-id 247310070 (Projects C02 and B06), GR\,4667/1. 
We acknowledge the CoreLab Quantum Materials, Helmholtz Zentrum Berlin für Materialien und Energie (HZB), Germany, where the single crystal samples of PbCuTe$_2$O$_6$ were synthesized and characterised.

\bibliography{PbCuTe}
\newpage

\end{document}